\documentclass{sig-alternate-05-2015}

\usepackage{graphicx}
\usepackage{algorithm}
\usepackage{algorithmic}
\usepackage{color}
\usepackage{times}
\usepackage{amsmath}
\usepackage{enumitem}
\usepackage{graphicx}
\usepackage{subcaption}

\newcommand{\eatproofs}[1]{#1}

\newcommand{\eatTR}[1]{}

\begin{document}
	
	
	\pdfoutput=1
	
	\title{Authority-Based Team Discovery in Social Networks }
	
	\author{
		\alignauthor
		Morteza Zihayat{\small $^{*}$}, Aijun An{\small $^{\$}$}, Lukasz Golab{\small $^\dag$}, Mehdi Kargar{\small $^\ddag$}, Jaroslaw Szlichta{\small $^{\#}$}\\
		\affaddr{$^{*}$University of Toronto, Toronto, Canada},
		\affaddr{$^{\$}$York University, Toronto, Canada}\\
		\affaddr{$^\dag$University of Waterloo, Waterloo, Canada},
		\affaddr{$^\ddag$University of Windsor, Windsor, Canada}\\              
		\affaddr{$^{\#}$University of Ontario Institute of Technology, Oshawa, Canada}\\
		\affaddr{mori.zihayatkermani@utoronto.ca, aan@cse.yorku.ca, lgolab@uwaterloo.ca,\\ mkargar@uwindsor.ca, jaroslaw.szlichta@uoit.ca}
	}

	\maketitle
	
	\newdef{definition}{Definition}
	\newdef{problem}{Problem}
	\newtheorem{theorem}{Theorem}
	
	\newcommand{\eat}[1]{}

	\begin{abstract}
Given a social network of experts, we address the problem of discovering a team of experts that collectively holds a set of skills required to complete a given project.  Most prior work ranks possible solutions by communication cost, represented by edge weights in the expert network.  Our contribution is to take experts' authority into account, represented by node weights.  We formulate several problems that combine communication cost and authority, we prove that they are NP-hard, and we propose and experimentally evaluate greedy algorithms to solve them.
		
	\end{abstract}
	

	\section{Introduction}
	
An expert network is a social network containing professionals who provide specialized skills or services.  Expert network providers include the employment-oriented service LinkedIn, the repository hosting service GitHub, and bibliography-based Websites such as DBLP and Google Scholar.  A node in an expert network corresponds to a person and node labels denote his or her areas of expertise.  Experts may be connected if they have previously worked together, co-authored a paper, etc.  Edge weights may denote the strength of a relationship, the number of co-authored publications, or the \emph{communication cost} between experts \cite{Li2015,Roy2015}.

There has been recent interest in the problem of finding teams of experts from such networks; see, e.g, \cite{Lappas2009,Roy2015}.  A common approach has been to find a subgraph of the expert network whose nodes collectively contain a given set of skills and whose communication cost is minimal.  In this paper, we argue that in many practical applications, other factors should also be considered.  For example, experts may be associated with {\em authority} metric such as h-index or number of publications. Here, we may want to minimize communication costs and maximize authority.  Furthermore, in large social networks, experts holding the desired skills may not be directly connected.  Thus, we may obtain a subgraph with some nodes, the \emph{skill holders}, corresponding to team members who have the desired skills, and other nodes serving as connectors.  The authority of connectors may also affect the quality of the team; e.g., \emph{connectors} may serve as mentors for the skill holders.  

\begin{figure}[t]
	\centering
	\includegraphics[width=8.0cm]{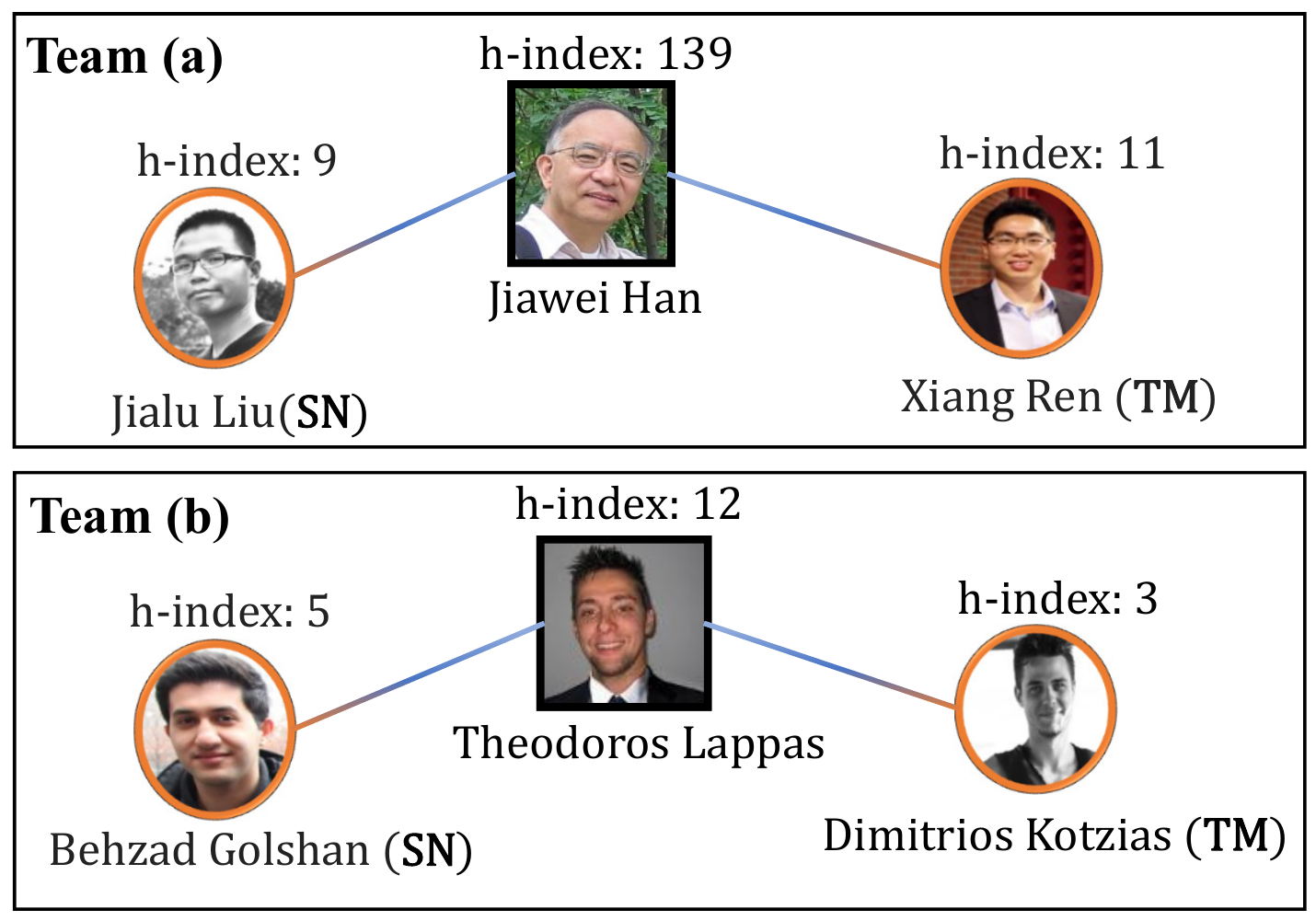}
	
	\caption{Two teams with expertise in SN and TM.} \label{fig:motivation}

\end{figure}

For instance, consider the two teams of researchers in Figure~\ref{fig:motivation}, both having expertise in social networks  (\emph{SN}) and text mining (\emph{TM}). Team (a) and (b) both have two skill holders and a connector node; in this example, we use graduate students as skill holders and professors as connectors. Assuming equal communication costs, i.e., each edge having the same weight, previous work cannot distinguish between these two teams. However, one can observe that the experts in team (a) have more experience (higher h-index).  Furthermore, even if all the skill holders were to have the same authority, team (a) may be preferable because its connector has more authority.

Our contributions are as follows.
\begin{enumerate}
	
	\item We formally define the problem of authority-based team formation in expert networks. We formulate three ranking objectives which optimize communication cost, skill holder authority, connector authority and combinations of them. We prove that optimizing these objectives is NP-hard.
	
	\item Since these problems are NP-hard, we propose greedy algorithms to solve them. We present an algorithm to optimize communication cost over an expert network $G$.  We then give a transformation which moves authority (node weights) onto the edges of a new graph, $G'$, and prove that our algorithm also optimizes the other objectives over $G'$. 
	
	\item 
	We perform a comprehensive evaluation using the DBLP dataset to confirm the effectiveness and efficiency of our approach.  In particular, we show that the teams discovered by our techniques perform higher-quality research than those found using prior work.
		
	
\end{enumerate}

\section{Preliminaries}
	
	Let $C = \{c_1, c_2, \dots , c_m \}$ be a set of $m$ experts, and $S = \{s_1, s_2, \dots , s_r\}$ be a set of $r$ skills.  An expert $c_i$ has a set of skills, denoted as $S(c_i)$, and $S(c_i)\subseteq S$.  If $s_j \in S(c_i)$, expert $c_i$ has skill $s_j$.   Furthermore, a subset of experts $C' \subseteq C$ have skill $s_j$ if at least one of them has $s_j$.  For each skill $s_j$, the set of all experts having skill $s_j$ is denoted as $C(s_j) = \{c_i | s_j \in S(c_i)\}$.  A project $P \subseteq S$ is a set of required skills.  A subset of experts $C' \subseteq C$  {\em covers} a project $P$ if $\forall s_j \in P \ \exists \ c_i \in C', s_j \in S(c_i)$.
	
	We model the social network of experts as an undirected graph $G$.  Each node in $G$ is an expert in $C$ (we use the terms expert and node interchangeably).  Each expert $c_i$ has an application-dependent authority   
	$a(c_i)$.  To convert authority maximization into a minimization problem, we set 
	$a'(c_i) = \frac{1}{a(c_i)}$.  Furthermore, let $w(c_i, c_j)$ be the weight of the edge between two experts $c_i$ and $c_j$.  Edge weights correspond to application-dependent communication cost or relationship strength. There is no edge between experts who have no relationship or prior collaboration.  Formally:

	\begin{definition}
		
		\textbf{Team of Experts:} Given an expert network $G$ and a project $P$ that requires the set of skills $\{s_1, s_2, \dots,$ $s_n\}$, a {\em team of experts} $T$ is a connected subgraph of $G$ whose nodes cover $P$.  With each team, we associate a set of $n$ skill-expert pairs: $\{\langle s_1, c_{s_1}\rangle, \langle s_2, c_{s_2}\rangle, \dots, \langle s_n, c_{s_n}\rangle\}$, where $c_{s_j}$ is an expert in $T$ that has skill $s_j$ for $j=1,\dots, n$. 
		
		\label{definition-team}
	\end{definition}

	The same expert may cover more than one required skill, i.e., $c_{s_i}$ can be the same as $c_{s_j}$ for $i\neq j$.  Also, there may not be a direct edge between some two experts $c_{s_i}$ and $c_{s_j}$ in $G$.  Thus, $T$ may include connector nodes that may not hold any skill in $P$ (e.g., Han and Lappas in Figure~\ref{fig:motivation}).  Assuming that edge weights denote communication costs, minimizing communication costs amounts to minimizing the sum of the weights of the team's edges \cite{Lappas2009}.	
	
	\begin{definition}
		\textbf{Communication Cost (CC):} Suppose the edges of a team $T$ are denoted as $\{e_1, e_2, \dots, e_t\}$.  The communication cost of $T$ is defined as CC($T$) $= \sum_{i=1}^{t} w(e_i)$, where $w(e_i)$ is the weight of edge $e_i$.
		\label{definition-team}
	
	\end{definition}

	\begin{problem}
		Given a graph $G$ and a project $P$, find a team of experts $T$ for $P$ with minimal communication cost CC($T$).
		\label{problem-comcost}
	
	\end{problem}
	
	This is an NP-hard problem \cite{Lappas2009} which has been studied before.  
	Extensions of this problem have also been considered, e.g., optimizing personnel cost and proficiency of skill holders
	\cite{kargar2013,zihayat2014}, or recommending replacements when a team member becomes unavailable \cite{Li2015}.  However, to the best of our knowledge, existing approaches do not optimize  the authority of skill holders and connectors.
	
	\begin{figure*}[t]
		\centering
		\includegraphics[width=9cm]{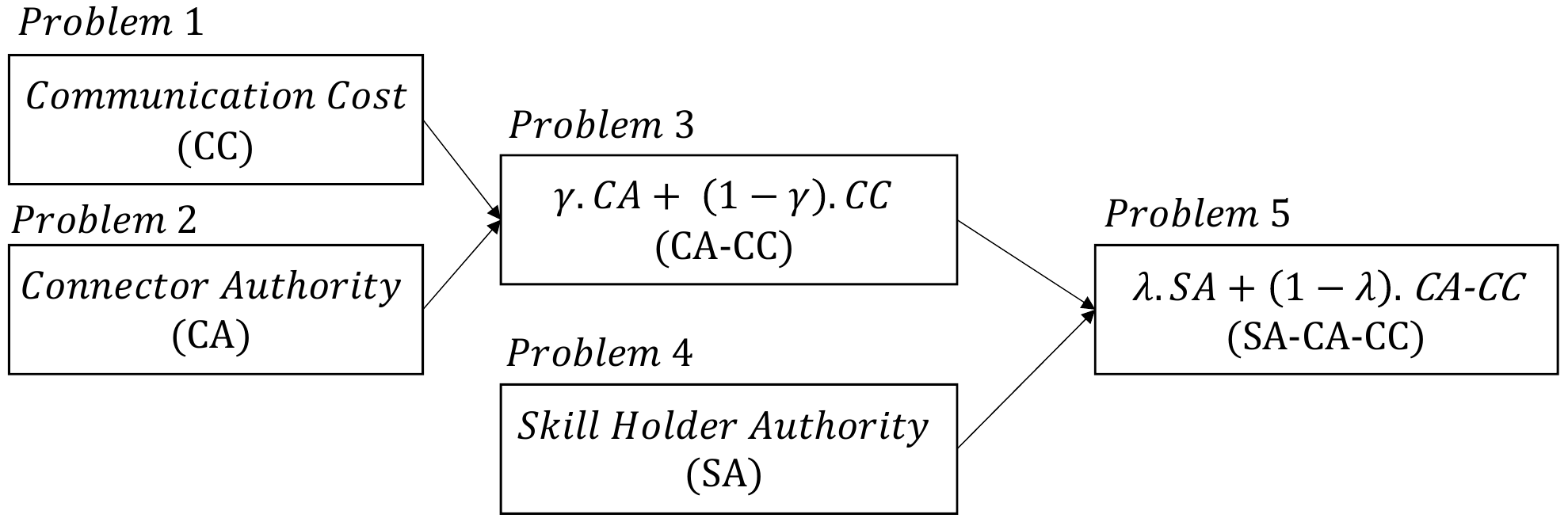}
	
		\caption{The Proposed Approach} \label{fig:Problems}

	\end{figure*}

	\section{Team Formation Framework}
	
	\subsection{Foundations}
	
	We are interested in optimizing both communication cost and authority.  Note that we optimize the authority of connectors and skill holders separately.  Some applications may find the authority of skill holders more important than that of the connectors (and vice versa), e.g., those where skill holders execute the project and connectors only provide guidance.  Therefore, we optimize them with different tradeoff parameters, $\gamma$ and $\lambda$, with respect to the communication cost and to each other.  Figure~\ref{fig:Problems} summarizes the problems we tackle and the remainder of this section discusses them in detail. First, we define the \emph{connector authority} of a team as the sum of the inverse-authorities $a'(c_i)$ of its connectors.
	
	\begin{definition}
		\textbf{Connector Authority (CA):} Suppose that the connectors of a team $T$ (all nodes excluding skill holders) are denoted as $\{c_1, c_2, \dots, c_q\}$.  The connector authority of $T$ is defined as CA($T$) $= \sum_{i=1}^{q} a'(c_i)$. 
		\label{definition-team}
	\end{definition}

	\begin{problem}
		Given a graph $G$ and a project $P$, find a team of experts $T$ for $P$ with minimal connector authority CA($T$).
		\label{problem-refauthority}
	\end{problem}
	
	\begin{theorem}
		Problem ~\ref{problem-refauthority} is NP-hard.
	\end{theorem}

	\eatproofs{	\begin{proof}

			We prove that the decision version of the problem is NP-hard. Thus, as a direct result, minimizing referral authority objective is NP-hard too.  The decision problem is specified as follows.  Given a graph $G$ and a set of required skills, determine whether there exists a team of experts with referral authority value of $const_{ra}$, for some constant $const_{ra}$.
			
			The problem is obviously in NP.  We prove the theorem by a reduction from group Steiner tree problem.  First, consider a graph in which all edges have the same weight of 1.0 and all nodes have the same authority of 1.0.  A feasible solution to the above problem with the referral authority at most $const_{ra}$ is a solution for the group Steiner tree problem with the weight at most ($const_{ra}$ - 1).  This is the case since for any tree, the number of edges is equal to the number of nodes minus one. Thus, if there exists a tree with the referral authority at most $const_{ra}$, then there exists a tree with the sum of the edge weights at most ($const_{ra}$ - 1).  On the other hand, a tree with edge weights at most ($const_{ra}$ - 1) determines a feasible tree with the referral authority at most $const_{ra}$. Therefore, the proof is complete.
		\end{proof}}

		\eatTR{Due to space limitations, we refer the reader to the extended version of this paper (technical report) for all proofs \cite{Zihayat2017}.}
		%
		%
		Furthermore, we are interested in the bi-criteria optimization problem of minimizing $CC$ and $CA$. To do so, we combine these two objectives into one with a tradeoff parameter $\gamma$ (after normalizing edge and node weights since they may have different scales).

		\begin{definition}
			
			\textbf{CA-CC Objective:}
			Given a team $T$ and a tradeoff parameter $\gamma$, where $0 \leq \gamma \leq 1$, the CA-CC score of $T$ is defined as CA-CC($T$) $= \gamma \times $CA$(T) + (1-\gamma) \times $CC$(T)$.
			\label{definition-firstCombObj}
		\end{definition}

		\begin{problem}
			Given a graph $G$, a project $P$, and a tradeoff parameter $\gamma$, find a team of experts $T$ for $P$ with minimal CA-CC($T$).
			\label{problem-fristcomb}
		\end{problem}
		
		\begin{theorem}
			Problem \ref{problem-fristcomb} is NP-hard.
			\label{theorem-fristcomb}
		
		\end{theorem}

		\eatproofs{	\begin{proof}
				
				We showed that finding a team of experts covering the input skills with minimized communication cost ($CC(T)$) or minimized referral authority ($RA(T)$) is NP-hard.  Since both $CC(T)$ and $RA(T)$ are linearly related to {\em CA-CC}(T) (the objective of Problem \ref{problem-fristcomb}), then minimizing {\em CA-CC}(T) is also an NP-hard problem.
				
			\end{proof}
		}
		
		We are also interested in optimizing the authority of skill holders.

		\begin{definition}
			
			\textbf{Skill Holder Authority (SA):} Suppose that the skill holders of a team $T$ are denoted as $\{c_1, c_2, \dots, c_n\}$.  The skill holder authority of $T$ is defined as $SA(T) = \sum_{i=1}^{n} a'(c_i)$.
			\label{definition-team}
		\end{definition}

		\begin{problem}
			Given a graph $G$ and a project $P$, find a team of experts $T$ for $P$ with minimal skill holder authority SA($T$).
			\label{problem-expauthority}
		\end{problem}

		Problem \ref{problem-expauthority} can be solved in polynomial time: for each skill in $P$, we find an expert with the highest $a$ (lowest $a'$), and then produce a connected subgraph containing the selected experts.  However, this ignores communication cost and connectors' authority.  We now put all three objectives together.

		\begin{definition}
			
			\textbf{SA-CA-CC Objective:}
			Given a team $T$ and a tradeoff parameter $\lambda$, where $0 \leq \lambda \leq 1$, the SA-CA-CC objective of $T$ is defined as SA-CA-CC($T$) $= \lambda \times $SA$(T) + (1-\lambda) \times $CA-CC$(T)$.

			\label{definition-secondCombObj}
		\end{definition}

		\begin{problem}
			Given a graph $G$, a project $P$, and a tradeoff parameter $\lambda$, find a team of experts $T$ for $P$ with minimal SA-CA-CC($T$).
			\label{problem-secondcomb}
		\end{problem}

		\begin{theorem}
			Problem \ref{problem-secondcomb} is NP-hard.
			\label{theorem-secondcomb}

		\end{theorem}

		\eatproofs{
			\begin{proof}
				We showed that finding a team of experts covering the input skills with minimized {\em CA-CC} objective ({\em CA-CC}(T)) is NP-hard. Since {\em CA-CC}(T) is linearly related to {\em SA-CA-CC}(T) (the objective of Problem \ref{problem-secondcomb}), then minimizing {\em SA-CA-CC}(T) is also an NP-hard problem.	
			\end{proof}}

			Since the tradeoff parameters $\gamma$ and $\lambda$ are application-dependent, we leverage user and domain expert feedback to set and update them over time (see experiment in Figure~\ref{fig:ParameterSensitivity}). Incorporating user feedback is important for achieving high precision.

	\begin{algorithm}[t]
		\caption{Finding Best Team of Experts}
		\label{alg:bestTeam}
		\textbf{Input}: graph $G$ with $N$ nodes; project $P = \{s_1, s_2, \dots , s_t \}$; the set of experts that contains each skill $s_i$, $C(s_i)$, for $i=1, \dots, t$.\\
		\textbf{Output}: best team of experts
		
		\begin{algorithmic}[1]
			
			\STATE $leastTeamCost \leftarrow \infty$
			
			\STATE $bestTeam \leftarrow \emptyset$
			
			\FOR{$r \leftarrow 1$ \TO $N$}
			
			\STATE $root \leftarrow c_r$
			
			\STATE $teamCost \leftarrow 0$
			
			\STATE $team \leftarrow \emptyset$
			
			\STATE set the root of $team$ to $root$
			
			\FOR{$i \leftarrow 1$ \TO $t$}
			
			\STATE $minCost_i \leftarrow \min_{v \in C(s_i)} DIST(root, v)$
			
			\STATE $bestExpert \leftarrow \arg\min_{v \in C(s_i)} DIST(root, v)$
			
			
			\IF{$bestExpert \neq \emptyset$}
			
			\STATE $teamCost \leftarrow teamCost + minCost_i$
			
			\STATE $team.$add($bestExpert$)
			
			\ENDIF
			
			\ENDFOR
			
			\IF{size($team$) = $t$}
			
			\IF{$teamCost < leastTeamCost$}
			
			\STATE $leastTeamCost \leftarrow teamCost$
			
			\STATE $bestTeam \leftarrow team$
			
			\ENDIF
			
			\ENDIF
			
			\ENDFOR
			
			\RETURN $bestTeam$
			
		\end{algorithmic}
	\end{algorithm}

\subsection{Search Algorithms}	
	
	Since Problems \ref{problem-comcost}, \ref{problem-refauthority}, \ref{problem-fristcomb} and \ref{problem-secondcomb} are NP-hard, we propose efficient and effective greedy algorithms to optimize them in polynomial time. 
$\\	\\$
	\subsubsection{Optimizing CC}
	Algorithm~\ref{alg:bestTeam} returns a subtree of $G$ corresponding to a team with the optimized communication cost (sum of edge weights). The for-loop in line 3 considers each expert $c_r$ as a potential root node for the subtree ($c_r$ may end up being a skill holder or a connector). To build a tree around $c_r$, for each required skill $s_i$, we select the nearest skill holder, denoted $bestExpert$, that contains $s_i$ (lines 9-13; assume DIST($v_1$,$v_2$) finds the shortest path, i.e., the smallest sum of edge weights, between two nodes $v_1$, $v_2$). The method $add$ in line 13 connects the $bestExpert$ to the current team, meaning that any additional nodes along the path from the root to $bestExpert$ are also added. The tree with the lowest sum of edge weights is the best team (lines 14-17).  To find the shortest path between any two nodes in constant time, we use {\em distance labeling}, or {\em 2-hop cover} \cite{akiba2013}.  As a result, the complexity of Algorithm \ref{alg:bestTeam} is $O(N \times t \times |C_{max}|)$, where $|C_{max}|$ is the maximum size of the expert sets $C(s_i)$ for $1\leq i\leq t$.  The $N$ comes from the for-loop in line 3, the $t$ comes from the for-loop in line 8 and the $|C_{max}|$ is due to computing the shortest path to each expert in $C(s_i)$ in lines 9 and 10. For finding top-$k$ teams, we initialize a list $L$ of size $k$ for the output. The list $L$ is updated after each iteration of the loop and the new team is added to $L$ if its cost is smaller than the last team in $L$. The runtime complexity remains the same as the entire operation only needs an extra pass over $L$ in each iteration. 
	
	To solve the other problems, we transform the expert network $G$ by moving authority (node weights) onto the edge weights and then running Algorithm \ref{alg:bestTeam} on the transformed graph.
	
	\subsubsection{Optimizing CA-CC} For Problem ~\ref{problem-fristcomb}, we transform $G$ into $G'$ as follows.  Let the edge weight between nodes $c_i$ and $c_j$ in $G$ be $w(c_i,c_j)$.  In $G'$, we transform each edge weight to $w'(c_i,c_j) = \gamma(a'(c_i)+a'(c_j)) + 2 \times (1-\gamma)w(c_i,c_j)$.    The DIST function now finds the shortest paths by adding up the \emph{transformed} edge weights $w'$.  However, we only want to take connector authority into account, not skill-holder authority.  Therefore, in lines 9 and 10, we 
	replace $DIST(root,v)$ by $DIST(root,v)-\gamma a'(v)$; note that $v$ is always a skill holder.  If $root$ contains skill $s_i$, then $DIST$ is set to zero and skill $s_i$ is assigned to $root$.  With this modification, we claim that running Algorithm \ref{alg:bestTeam} on $G'$ optimizes CA-CC.
	Note that setting $\gamma=1$ solves Problem \ref{problem-refauthority}, i.e., optimizes CA.

\begin{figure*}[t]
	\centering
	\includegraphics[width=10cm]{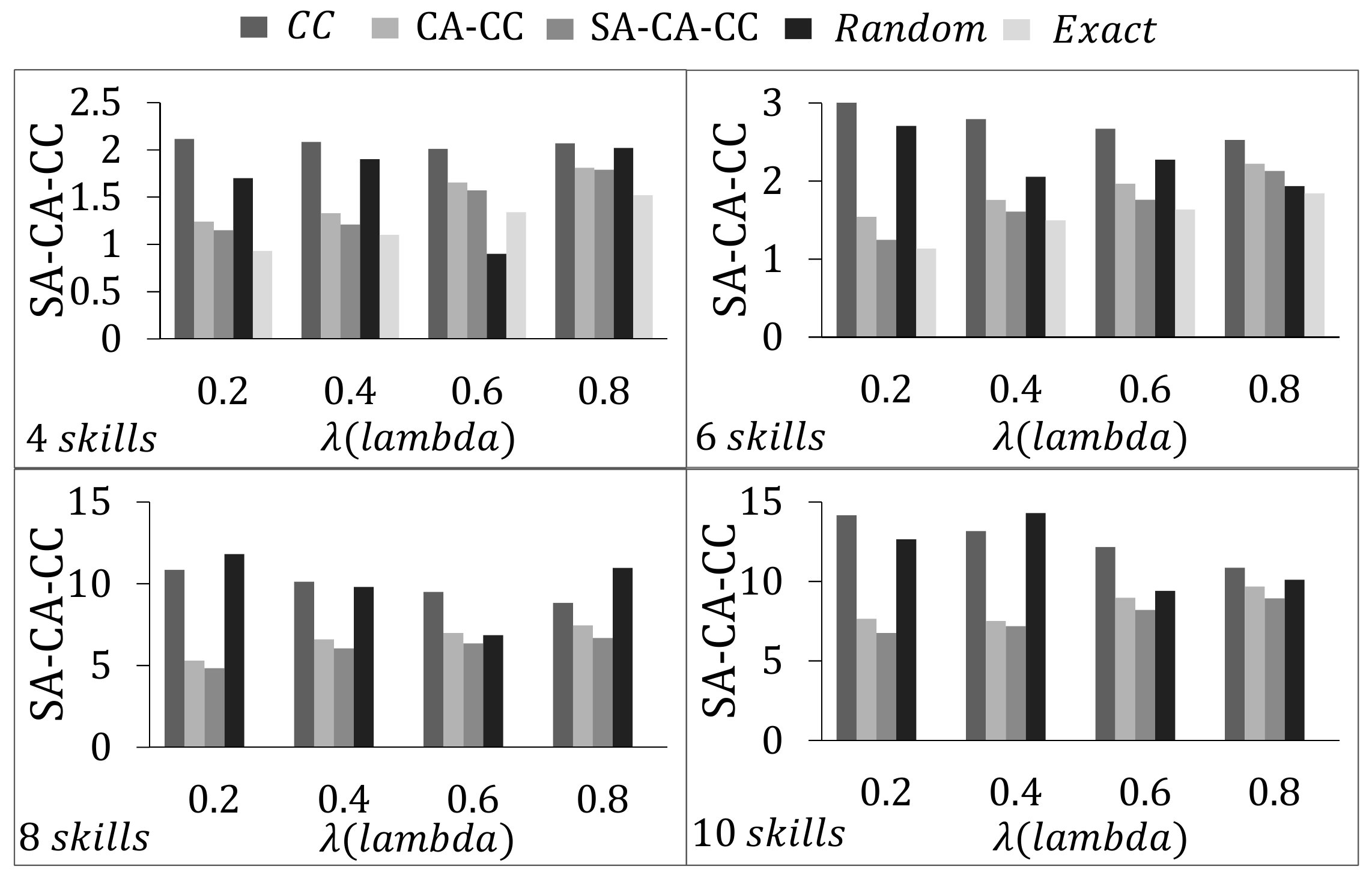}

	\caption{{\em SA-CA-CC} scores of different ranking methods($\gamma$ = $0.6$)} \label{fig:CombCostResults}

\end{figure*}
 \subsubsection { Optimizing SA-CA-CC}  
  Recall that SA-CA-CC is a linear combination of communication cost, skill holder authority and connector authority. We re-use $G'$ from above to capture communication cost and connector authority.  Additionally, we need to take $\lambda$ into account and add the contribution of skill holder authority.  To do this, we replace $DIST(root,v)$ in lines 9 and 10 with $(1-\lambda)(DIST(root,v)-\gamma a'(v)) + \lambda a'(v)$. 
  Note that we have to subtract the authority of skill holders with parameter $\gamma$ and then add it with parameter $\lambda$. As before, if $root$ contains skill $s_i$, then $DIST$ is set to zero and skill $s_i$ is assigned to $root$. We claim that running Algorithm \ref{alg:bestTeam} with this modification, along with using $G'$ instead of $G$, solves Problem \ref{problem-secondcomb}.  

\section{Experimental Results}\label{section-experiments}
\begin{figure}[t]
	\centering
	\includegraphics[width=8cm]{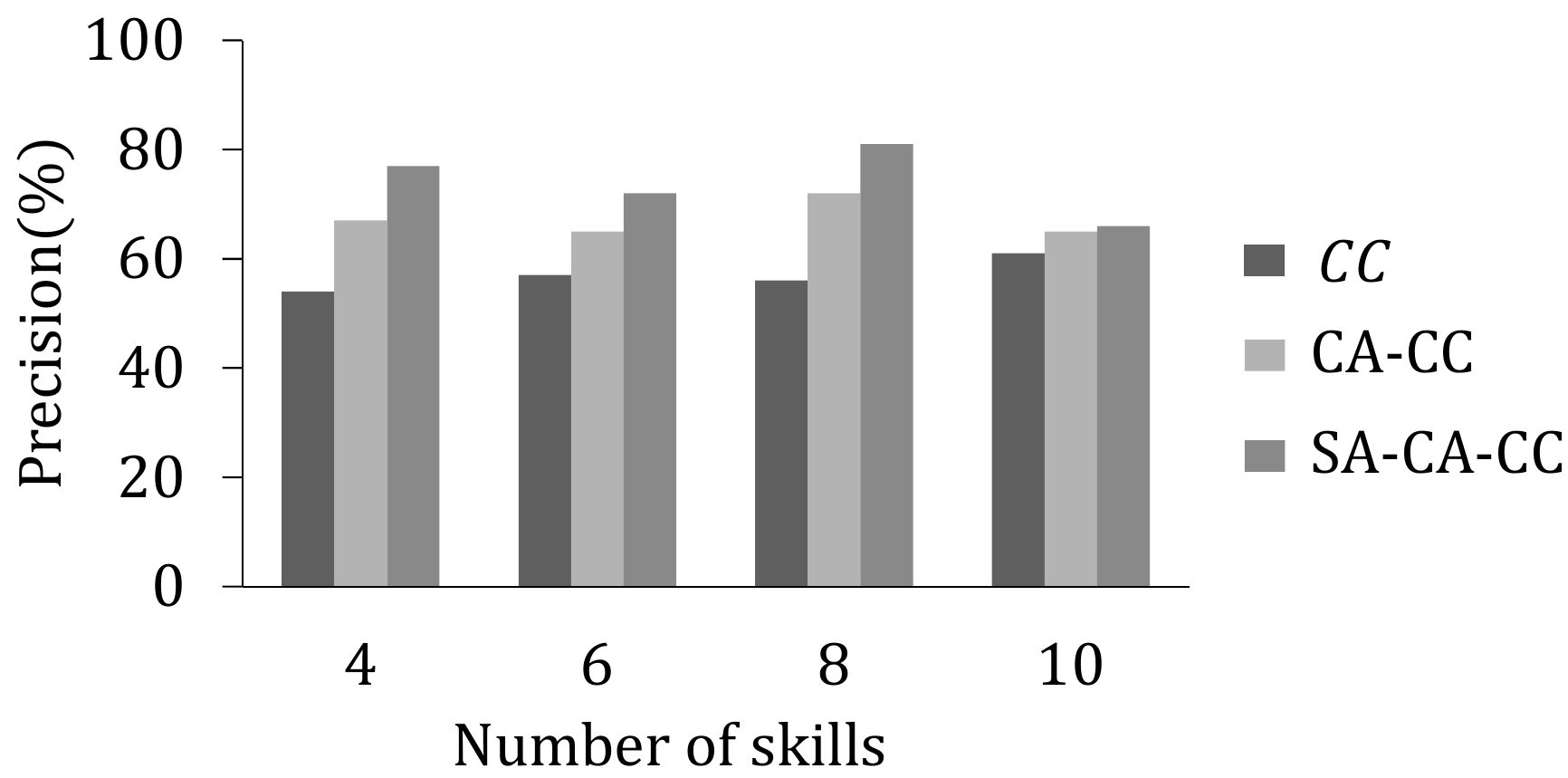}

	\caption{Precision of top-5 teams for different methods} \label{fig:precision}

\end{figure}
In this section, we use Algorithm~\ref{alg:bestTeam} and its various modifications explained above to implement ranking strategies for team discovery which optimize $CC$, {\em CA-CC} and {\em SA-CA-CC}.  $CC$ corresponds to prior state-of-the-art, and our main goal is to show that {\em CA-CC} and {\em SA-CA-CC} are more effective.  We also implemented $Random$, which randomly builds 10,000 teams and selects the one with the lowest {\em SA-CA-CC}, and $Exact$ which performs exhaustive search to find an ({\em SA-CA-CC})-optimal solution.  Note, however, that $Exact$ is intractable for large networks or large projects (containing many required skills).  The algorithms are implemented in Java and the experiments are conducted on an Intel(R) Core(TM) i7 2.80 GHz computer with 4 GB of RAM.

\begin{figure*}[t]
	\centering
	\includegraphics[width=10cm]{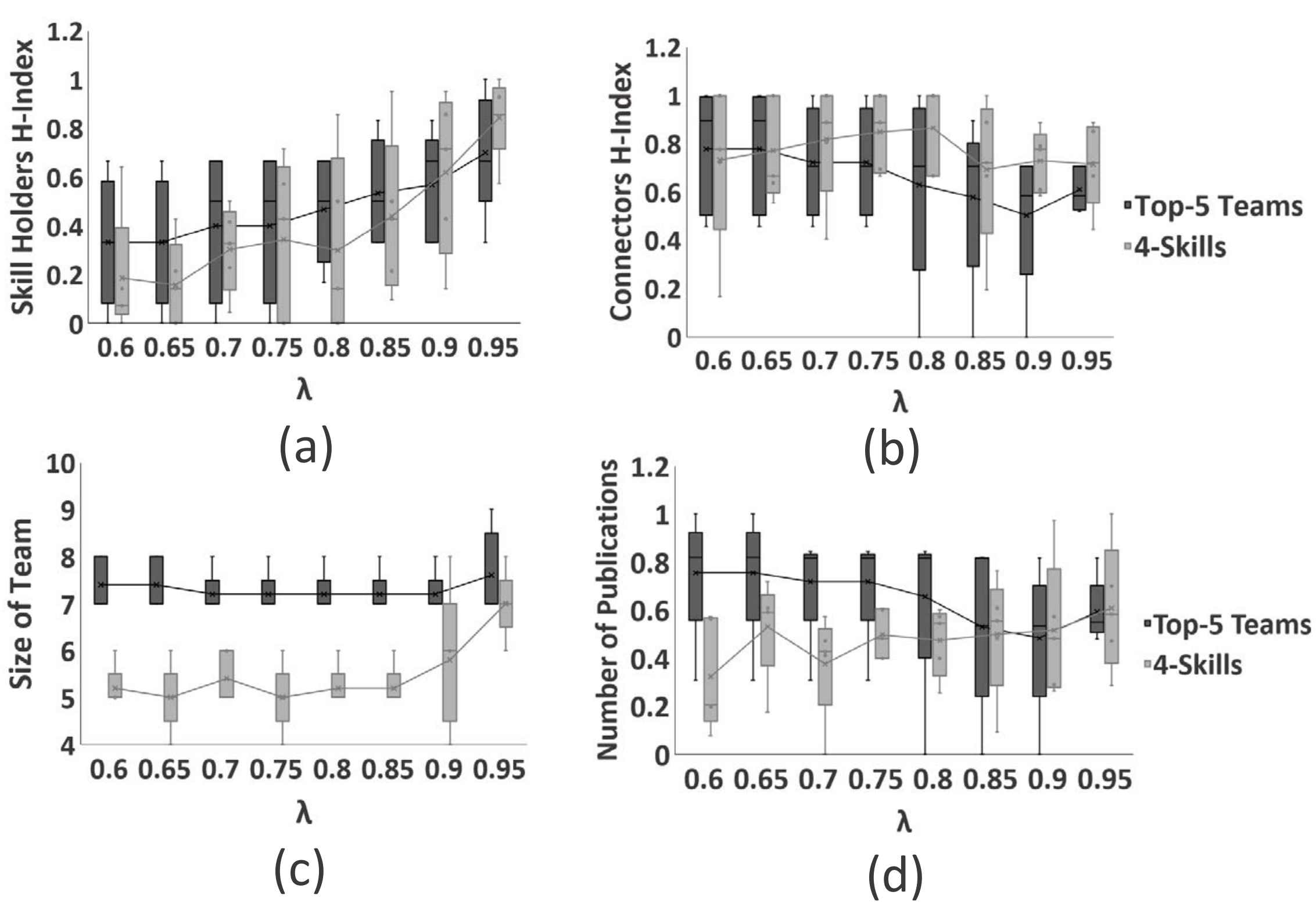}

	\caption{Sensitivity of normalized results to $\lambda$} \label{fig:ParameterSensitivity}
\vspace{0.2in}	
\end{figure*}
Similar to previous work, we use the DBLP XML dataset\footnote{http://dblp.uni-trier.de/xml/} to build an expert graph \cite{kargar2013,Lappas2009}.  For potential skill holders, we take junior researchers with fewer than 10 papers and we label them with terms that occur in at least two of their paper titles. This gives us the areas of expertise. Similar to \cite{kargar2013,Lappas2009}, we set edge weights between two experts $c_i$ and $c_j$ to $1 - |\frac{b_{c_i} \cap \ b_{c_j}}{b_{c_i} \cup \ b_{c_j}}|$ (Jaccard Similarity) where $b_{c_i}$ is the set of papers of author $c_i$. 
We use h-index as the node weight to denote authority.  The resulting graph has 40K nodes (experts) and 125K edges.  The number of skills in a project is set to 4, 6, 8 or 10.  For each number of skills, we generate 50 sets of skills, corresponding to 50 projects, and we report average results over these 50 projects. 
\begin{figure*}[t]
	\centering

	\begin{subfigure}{.3\textwidth}
		\centering
		\includegraphics[width=1\linewidth,height=5cm]{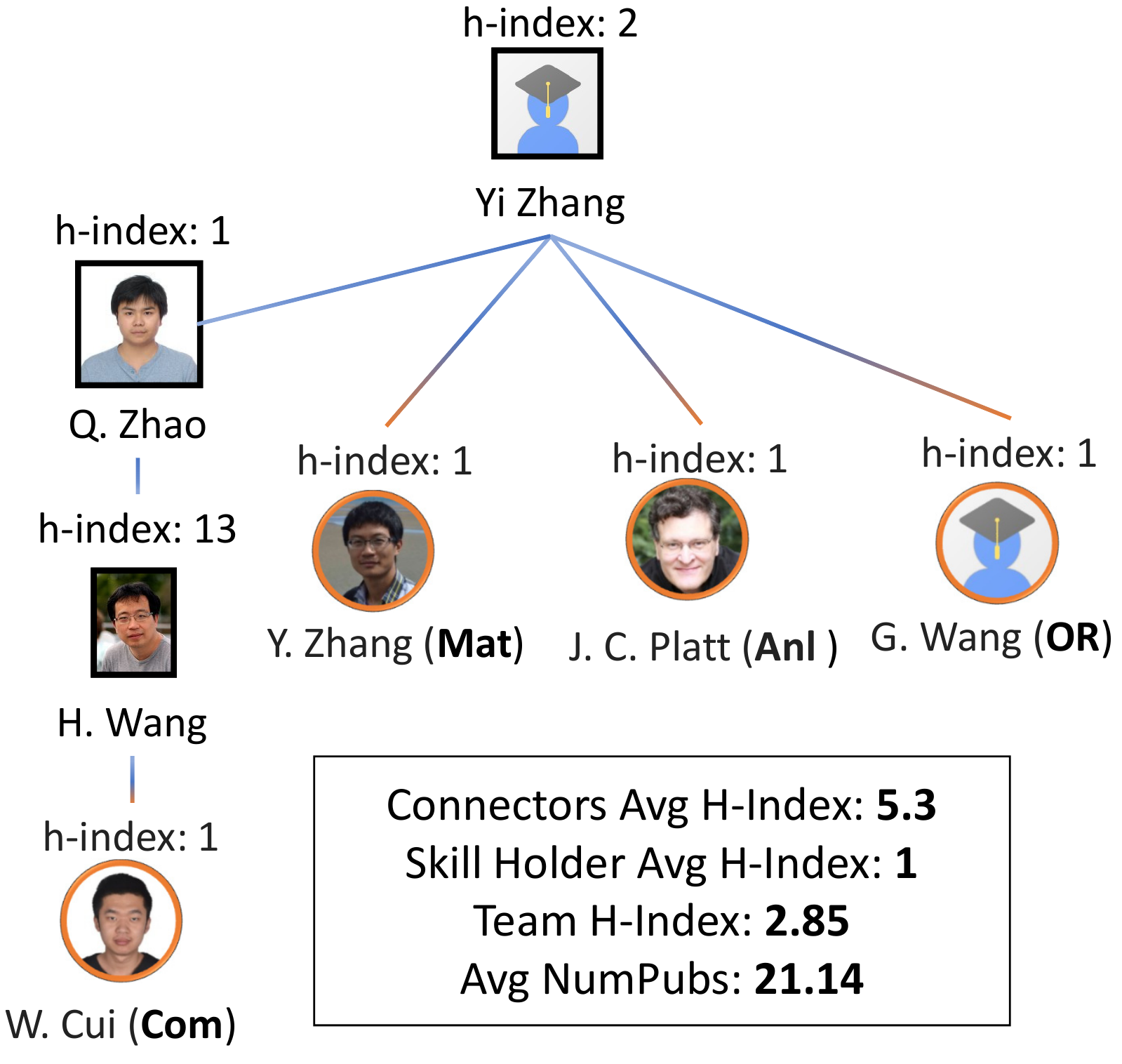}
		\caption{(CC)}
		\label{fig:sfig1}	
	\end{subfigure}%
	\begin{subfigure}{.32\textwidth}
		\centering
		\includegraphics[width=1\linewidth,height=5cm]{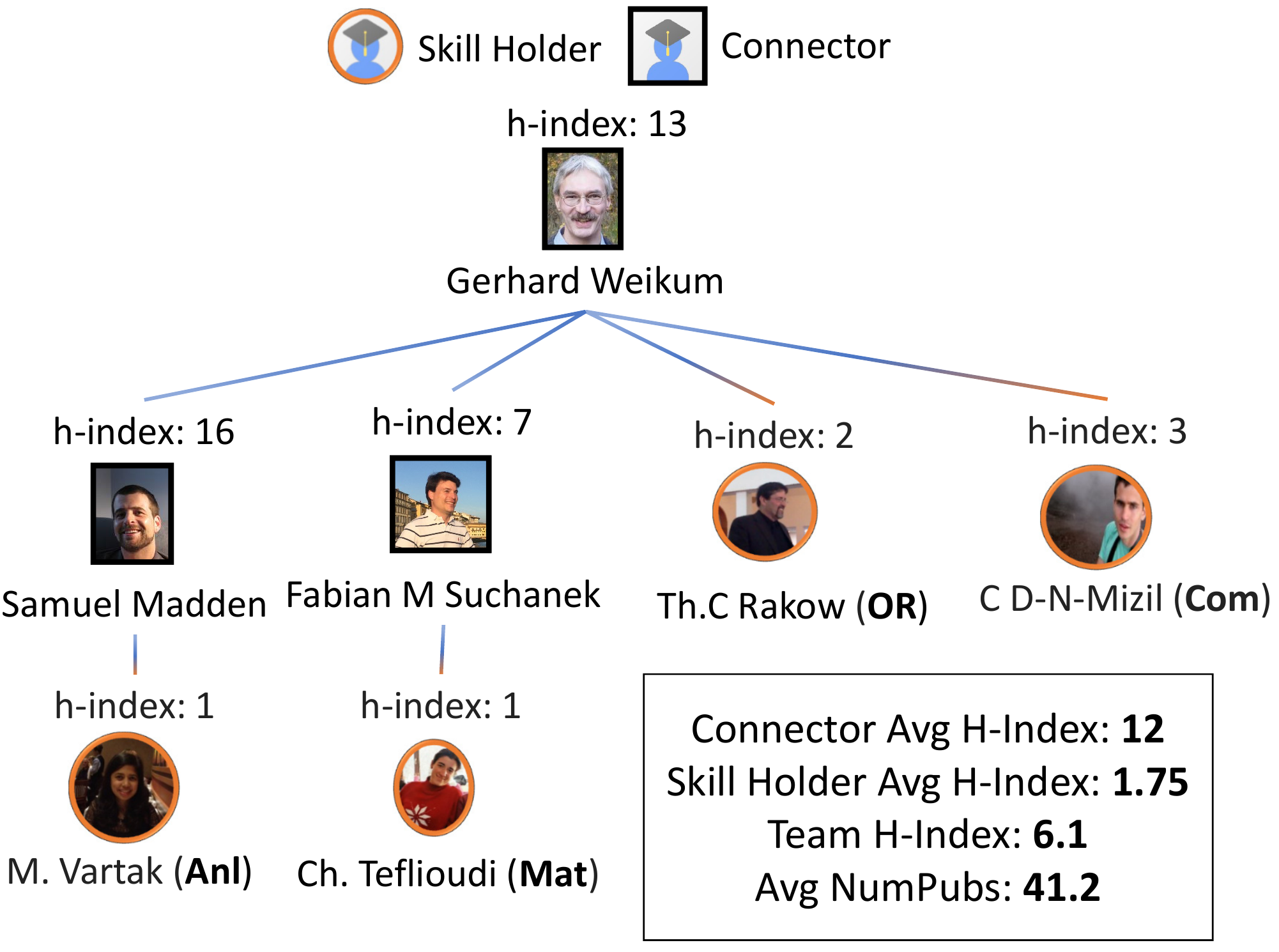}
		\caption{(CA-CC)}
		\label{fig:sfig2}
	\end{subfigure}	
	\begin{subfigure}{.35\textwidth}
		\centering
		\includegraphics[width=1\linewidth,height=5cm]{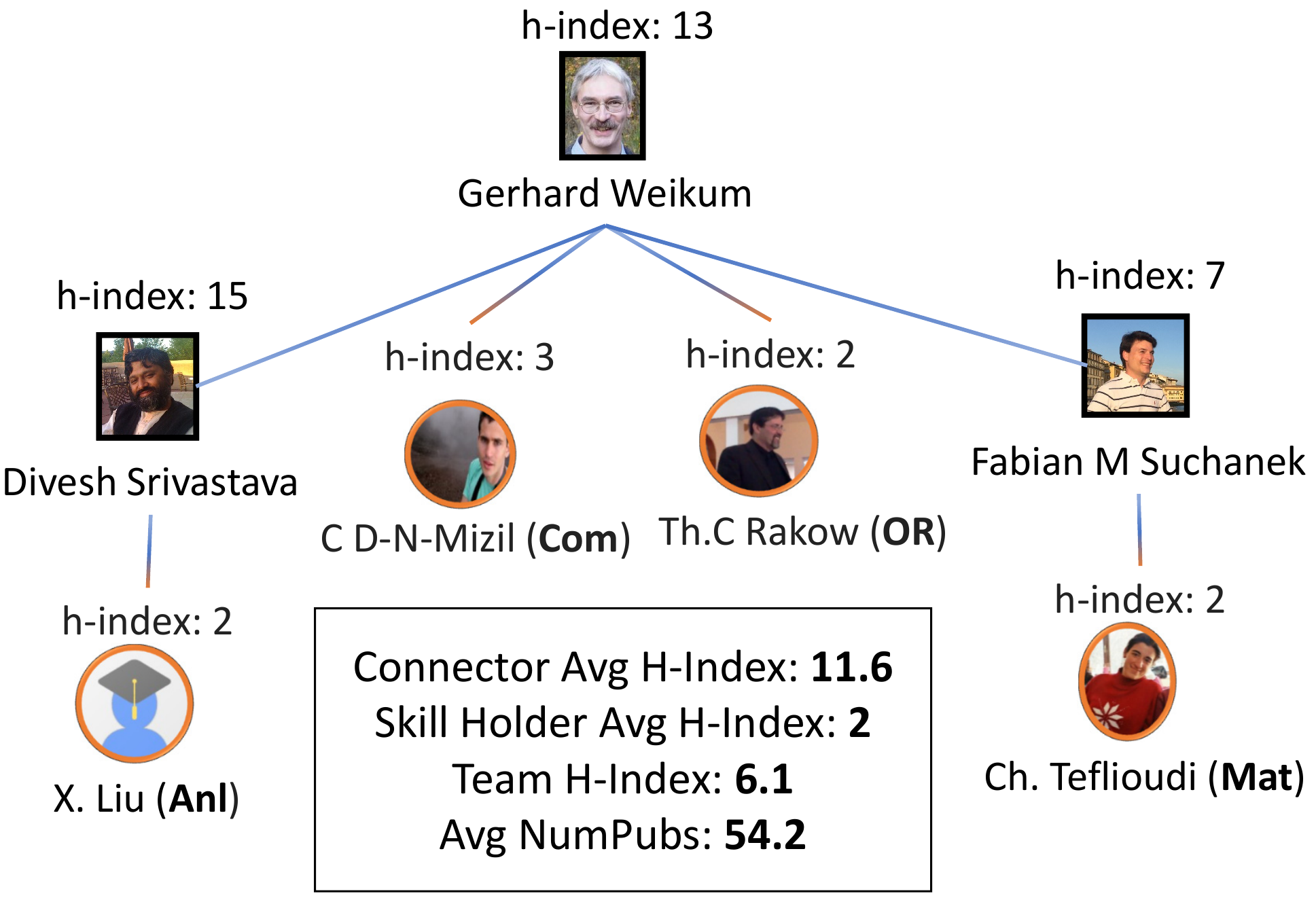}
		\caption{(SA-CA-CC)}
		\label{fig:sfig2}
	\end{subfigure}	
	\caption{Best team of $CC$, {\em CA-CC} and {\em SA-CA-CC} with "skills": {\em analytics({\bf Anl}), matrix ({\bf Mat}), communities({\bf Com}), object oriented({\bf OR})} } \label{fig:BestTeam}
	
\end{figure*}
$\\$
\subsection{Effectiveness}
We begin by comparing our {\em SA-CA-CC} ranking strategy with $Exact$; for completeness, we also test $CC$, {\em CA-CC} and $Random$, and compute their {\em SA-CA-CC} scores.  Figure \ref{fig:CombCostResults} plots the {\em SA-CA-CC} scores of different ranking strategies for different numbers of skills and different values of $\lambda$.  For brevity, we fix $\gamma$ at 0.6 but different values led to similar conclusions.  We conclude that 
{\em SA-CA-CC} produces results that are close to those of $Exact$ (but note that $Exact$ was only able to handle 4 and 6 skills and did not terminate in reasonable time for 8 and 10 skills).  Not surprisingly, {\em SA-CA-CC} has lower {\em SA-CA-CC} score than $CC$ and {\em CA-CC}.  We also note $CC$, {\em CA-CC} and {\em SA-CA-CC} have similar runtime since they use the same fundamental algorithm and indexing methods. The runtime depends on the number of required skills and is around a few hundred milliseconds (i.e., less than one second) on average.

\subsection{User Study}
 We conduct a user study to evaluate the top-$k$ precision of different ranking strategies.  First, we create four projects with different numbers of required skills.  Then, for each project, we run $CC$, {\em CA-CC} and {\em SA-CA-CC} and take the top-5 best teams returned by each.  We give these results to six Computer Science graduate students, along with the average number of publications and the h-index of each expert included in the teams.  We asked the students to judge the quality of the top-5 teams using a score between zero and one. Figure \ref{fig:precision} shows the top-5 precision of each method.  In this experiment, we set both $\lambda$ and $\gamma$ to 0.6. Both of our methods, {\em CA-CC} and {\em SA-CA-CC}, obtain better precision than $CC$ for all tested projects.

\subsection{Quality of Teams}
We check if the top-5 teams returned by {\em CC} and {\em SA-CA-CC} were successful in real life. To do so, we examined the rankings of the publication venues of these teams according to the {\em Microsoft Academic} conference ranking.  Since we used the {\em DBLP} dataset up to 2015 for team discovery, we only consider papers published in 2016. We set $\gamma$ and $\lambda$ to 0.6 and generate 5 different projects with four different skills. From the teams that co-authored papers in 2016, we found that 78\% of the time the teams found by {\em SA-CA-CC} published in more highly-rated venues than those found by CC.

\subsection{Sensitivity}
Figure \ref{fig:ParameterSensitivity} shows the sensitivity of the results to $\lambda$ (the tradeoff parameter between skill holder authority and {\em CA-CC}), specifically the sensitivity of the average h-index of skill holders (part a), the average h-index of connector nodes (part b), the average team size (part c) and the average number of publications (part d).  Our methodology for evaluating sensitivity is as follows.    
First, we examine the effect of $\lambda$ on the top 5 teams returned by {\em SA-CA-CC}.  Given the project [\emph{analytics, matrix, communities, object  oriented}], {\em SA-CA-CC} finds top-5 teams using different values of $\lambda$. Second, we evaluate the effect of $\lambda$ on a best team returned by {\em SA-CA-CC} for $m$ different projects.  For this, we randomly generate five projects with four skills each.  Then, for each value of $\lambda$, {\em SA-CA-CC} finds the best team for each project.  As shown in Figure \ref{fig:ParameterSensitivity}, the measures change slowly as $\lambda$ increases.  We also observe that changing the value of $\lambda$ by less than $0.05$ does not affect the results and the quality of the team remains the same. 

\subsection{Qualitative Evaluation}
 Figure~\ref{fig:BestTeam} illustrates the teams returned by $CC$, {\em CA-CC} and {\em SA-CA-CC} for the project [\emph{analytics, matrix, communities, object oriented}]. Observe that $CC$ returns a team with lower authority (average h-index) and average number of publications than {\em CA-CC} and {\em SA-CA-CC}.  Moreover, Figure \ref{fig:BestTeam} shows that the skill holders of the team returned by {\em CA-CC} and {\em SA-CA-CC} are connected through authors with a higher h-index, and thus have a higher referral authority. We argue that the teams returned by our algorithms are more effective than the one returned by $CC$ since it reveals a deeper connection among the experts that may not have been discovered by existing team formation methods. Note that connectors may not be directly involved in performing a task, but may provide guidelines and support to skill holders.

\section{Conclusions} 
In this paper, we studied the problem of team discovery from networks of experts. We formulated new ranking objectives that take communication costs among experts as well as expert authority into account.  We proved that satisfying these new objectives is NP-hard and proposed heuristic algorithms.  We demonstrated the effectiveness of our techniques on the DBLP dataset. Another way to jointly optimize the communication cost and
expert authority objectives is to find a set of Pareto-optimal
teams. In the future, we plan to develop algorithms to find
such teams and rank them based on relevant measures of
interestingness.

\bibliographystyle{abbrv}

\bibliography{biblioGraph}

\end{document}